\documentclass[a4paper,11pt]{article}
\usepackage[swedish,english]{babel}
\usepackage{a4wide}
\usepackage{authblk}
\usepackage{amsmath,amsthm}
\usepackage{amssymb}
\usepackage{algorithm}
\usepackage[noend]{algpseudocode}
\usepackage{dsfont}
\usepackage{subcaption}
\usepackage{sidecap}
\sidecaptionvpos{figure}{t}
\usepackage[pdftitle={Bayesian epidemiological modeling over high-resolution network
  data},
  pdfauthor={Engblom, Eriksson, Widgren},
  pdffitwindow=true,
  breaklinks=true,
  colorlinks=true,
  urlcolor=blue,
  linkcolor=red,
  citecolor=red,
  anchorcolor=red]{hyperref}
\usepackage[numbers,sort&compress]{natbib}
\usepackage{numprint}
\npthousandsep{,}
\usepackage{pdfpages}

\numberwithin{equation}{section}
\numberwithin{table}{section}
\numberwithin{figure}{section}

\renewcommand{\Pr}{\mathbf{P}}
\DeclareMathOperator{\Expect}{\mathbb{E}}

\newcommand{\Realdom}{\mathbf{R}}
\newcommand{\Intdom}{\mathbf{Z}}

\newcommand{\SimInf}{\textit{SimInf}}

\newcommand{\review}[1]{#1}

\usepackage{gensymb}
\usepackage{booktabs}
\usepackage{xr}
\DeclareMathOperator{\Prev}{Prevalence}
\DeclareMathOperator{\Swab}{Swab}

\newcommand{\figref}[1]{Fig.~\ref{#1}}

\newcommand{\tabref}[1]{Tab.~\ref{#1}}

\renewcommand{\algref}[1]{Alg.~\ref{#1}}

\title{Bayesian epidemiological modeling \\ over high-resolution
  network data}

\author[1]{Stefan Engblom\thanks{Corresponding author: S.~Engblom,
    telephone +46-18-471 27 54, fax +46-18-51 19 25.}}

\author[1]{Robin Eriksson}

\author[2]{Stefan Widgren}

\affil[1]{{\footnotesize Division of Scientific Computing, Department
    of Information Technology, Uppsala University, SE-751 05 Uppsala,
    Sweden. E-mails: \href{mailto:stefane@it.uu.se}{stefane@it.uu.se},
    \href{mailto:robin.eriksson@it.uu.se}{robin.eriksson@it.uu.se}.}}

\affil[2]{{\footnotesize Department of Disease Control and
    Epidemiology, National Veterinary Institute, SE-751 89 Uppsala,
    Sweden. E-mail:
    \href{mailto:stefan.widgren@sva.se}{stefan.widgren@sva.se}}}

\date{}

\begin{document}

\selectlanguage{english}

\maketitle

\begin{abstract}

  Mathematical epidemiological models have a broad use, including both
  qualitative and quantitative applications. With the increasing
  availability of data, large-scale \emph{quantitative} disease spread
  models can nowadays be formulated. Such models have a great
  potential, e.g., in risk assessments in public health. Their main
  challenge is model parameterization given surveillance data, a
  problem which often limits their practical usage.


  We offer a solution to this problem by developing a Bayesian
  methodology suitable to epidemiological models driven by network
  data. The greatest difficulty in obtaining a concentrated parameter
  posterior is the quality of surveillance data; disease measurements
  are often scarce and carry little information about the
  parameters. The often overlooked problem of the model's
  identifiability therefore needs to be addressed, and we do so using
  a hierarchy of increasingly realistic known truth experiments.


  Our proposed Bayesian approach performs convincingly across all our
  synthetic tests. From pathogen measurements of shiga toxin-producing
  \textit{Escherichia coli} O157 in Swedish cattle, we are able to
  produce an accurate statistical model of first-principles confronted
  with data. Within this model we explore the potential of a Bayesian
  public health framework by assessing the efficiency of disease
  detection and -intervention scenarios.

\end{abstract}

\bigskip
\noindent
\textbf{Keywords:} Bayesian parameter estimation,
Pathogen detection, Disease intervention, Synthetic Likelihood,
Spatial stochastic models.


%


\section{Introduction}

Mathematical and computational modeling are the dominating approaches
in the analysis of the dynamics of diseases. The latter approach
becomes increasingly important as the amount of relevant data
grows. Large-scale computational epidemiological models have been
successfully employed to evaluate and inform disease mitigation
strategies \cite{ferguson2003planning, ferguson2005strategies,
  germann2006mitigation, degli2008mitigation,
  merler2015spatiotemporal, Brooks-Pollock2014ABCSMC}. With the
increasing qualities of data, the possibility of enhancing the
resolution through data integration down to the scale of single
individuals in a large population has also been realized
\cite{eubank2004modelling, ferguson2005strategies,
  balcan2009multiscale, merler2011determinants,
  Brooks-Pollock2014ABCSMC}. In similar spirit, detailed contact data
has been used to drive models of disease spread at various population
sizes \cite{salathe2010high, stehle2011simulation,
  bajardi2012optimizing, obadia2015detailed, toth2015role,
  Brooks-Pollock2014ABCSMC}. Data-driven models have aided in an
understanding of epidemic outbreaks and endemic conditions on scales
and at a level of detail that were not previously possible
\cite{ferguson2005strategies, bajardi2012optimizing, zhang2017spread,
  liu2018measurability, siminf3, Brooks-Pollock2014ABCSMC}.

Many infectious human diseases have a zoonotic origin, e.g.,
salmonellosis or infection by shiga toxin-producing
\textit{Escherichia coli} O157 (STEC O157) \cite{vtec2008_09,
  vtec2011_12}. Taking a \emph{One Health} perspective, we argue that
to address challenges with existing and emerging threats of zoonotic
disease, the aim for computational animal disease models should be to
take their place as an integrated part in public health
evaluations. This clearly puts high demands on the accuracy and the
way any modeling uncertainties are handled. Indeed, an outstanding
difficulty with most disease spread models is parameterization, an
issue often solved via mosaic approaches \cite{ferguson2005strategies,
  merler2011determinants, siminf3, fournie2018dynamic,
  brouwer2018epidemiology}, that is, relying on a combination of
published parameters and residual minimization schemes conditioned on
the data at hand. \review{This typically leads} to parameter point
estimates which always leave some doubts on the model's explanatory
power. In this respect Bayesian modeling approaches
\cite{McKinley2018review, Brooks-Pollock2014ABCSMC} are clearly
favored through their ability to consistently address probabilistic
hypotheses.

The main contribution of this paper is a feasibility demonstration of
Bayesian parameterization in a first-principle and data-driven
national scale epidemiological model. Moreover, this is achieved under
realistic assumptions as to the available pathogen
measurements. \review{Specifically, we do not require detailed
  sampling of large parts of the underlying epidemiological state, but
  only rely on detection protocols at the single-node level}. For this
purpose we consider a general class of disease spread models governed
by two transmission processes: within-node spread coupled with a
between-node spread via a transportation network. Technically, the
Bayesian posterior exploration algorithm we develop is based on
bootstrapped synthetic likelihoods and an adaptive Markov-chain Monte
Carlo algorithm.

An often overlooked issue is the model's \emph{identifiability}. That
is, the fundamental possibility of accurately deducing the model's
parameters, either in the limit of increasing amounts of data, or more
practically, for the data that is actually available. We analyze this
experimentally through a hierarchy of increasingly realistic
data-synthetic experiments. In this way we monitor the successive
complications due to increasingly realistic observational
data. Arguably, in many applications the most challenging issue is the
sparsity of disease measurement data. The actual information content
of even relatively expensive measurements is shallow and tells little
about the model's dynamics. Despite this challenge, we are able to
demonstrate the feasibility of Bayesian parameterization using actual
measurements taken from a study of the spread of STEC O157 in cattle
\cite{vtec125}, and this data is sparse, noisy, and (seemingly at
least) weakly informative.

With our holistic Bayesian modeling approach, a clear improvement of
the public health's arsenal of tools related to zoonotic diseases is
possible. To highlight this the paper is concluded by putting this
idea to the test in realistic detection- and intervention scenarios.


\section{Methods}

Our methodology is fully simulation-based and consists of a stochastic
epidemiological model and an associated simulation engine confronted
with measurements via Bayesian methods driven by synthetic likelihoods
\cite{wood2010statistical}. To initially judge the model's
identifiability we first approach the parameterization using simpler
approximate Bayesian methods and synthetic data with an available
ground-truth. Upon success, full posterior exploration via synthetic
likelihoods and adaptive Metropolis sampling is attempted, yielding an
overall useful parameterized model. To critically assess the quality
of the inference procedure we employ a version of \emph{parametric
  bootstrap}, and the Bayesian model itself may of course be validated
against external sources whenever possible.

Below, we first summarize the epidemiological model and the data
driving the simulations in
\S\S\ref{subsec:model}--\ref{subsec:data}. The Bayesian methodology is
worked through in detail in \S\S\ref{subsec:ABC}--\ref{subsec:error}.

\subsection{The epidemiological model}
\label{subsec:model}

\begin{figure}[t]
  \centering
  \includegraphics[width=1\linewidth, angle=0]{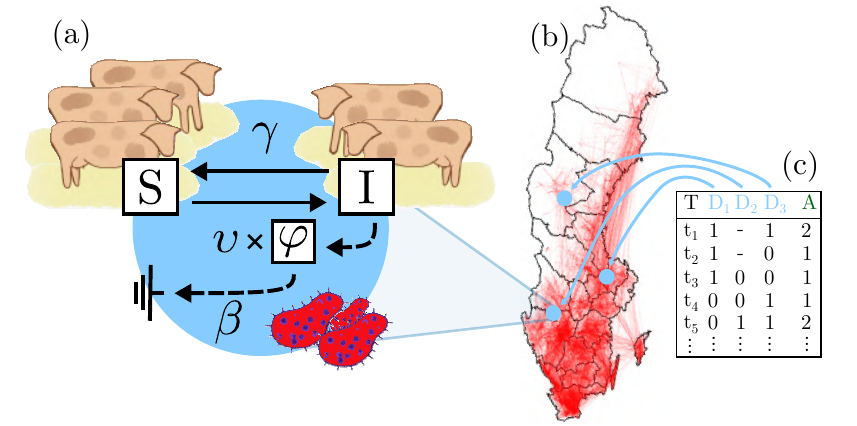} 
  \caption{{\bf (a)} The SIS$_E$ three parameter model. Susceptible
    individuals $S$ turn infected at rate $\upsilon \times
    \varphi$. Infected individuals $I$ recover at a rate $\gamma$, and
    shed the pathogen to the environment adding to the infectious
    pressure $\varphi$, which decays at rate $\beta$. {\bf (b)}
    National transport data form the dynamic contact network in the
    simulations (Sweden). {\bf (c)} Time series data $D_1$--$D_3$ from
    single node measurements (0: not detected pathogen, 1: detected
    pathogen, -: missing measurement). This data is summed up and the
    aggregated series $A$ is passed to the Bayesian inference
    procedure.}
  \label{fig:modelIntro}
\end{figure}

In our computations we used the \SimInf\ epidemiological engine
\cite{siminf_manual}, which allows completely general disease spread
models to be formulated. However, with the specifics of the STEC O157
application in mind, the model we have come to favor is the SIS$_E$
model \cite[Chap.~11]{FreeLivingInfectiveStages}, which contains three
state variables, $[S,I,\varphi]$, and which is defined firstly by the
Markovian transitions between the integer (counting) compartments,
\begin{align}
  \label{eq:SIS}
  S &\xrightarrow{\upsilon \varphi} I, \qquad
  I \xrightarrow{\gamma} S,
\end{align}
that is, susceptible individuals turn infected at a rate proportional
to the concentration of infectious matter $\varphi$, and infected
individuals recover at rate $\gamma$. Secondly, the environmental
variable \review{$\varphi$ represents the local environmental pathogen
  load and} obeys the non-dimensionalized \review{deterministic}
dynamics \cite{siminf_Ch}
\begin{align}
  \label{eq:_E}
  \varphi'(t) &= I (S+I)^{-1} -\beta \varphi,
\end{align}
in which infected individuals shed infectious matter into the
environment, which decays at a fixed rate $\beta$. While the basic
SIS$_E$ model can certainly be extended in various ways, we have found
that, in practice, \eqref{eq:SIS}--\eqref{eq:_E} balance model
complexity with typically available data quite well. The SIS$_E$ model
is schematically summarized in \figref{fig:modelIntro}~(a).

Using \SimInf, we connect local copies of the SIS$_E$ model with
transports of individuals over a dynamic \review{network, thus
  forming our} population of interest. These movements are either
synthetically generated or are pre-recorded movements from an actual
transport network, cf.~\figref{fig:modelIntro}~(b).

In summary, our epidemiological model consists of a three-parameter
local SIS$_E$ model in the form of a continuous-time Markov chain for
$(S,I)$, and an ordinary differential equation for $\varphi$,
connected over a network of nodes using \review{recorded} transport
data \review{at the level of batches of animals}. For the specific
STEC O157 application, in order to better fit data collected from a
country with fairly large climate variations (Sweden), the decay
parameter $\beta$ was separated into seasons $(\beta_{1,2,3,4})$ for
$[$spring,summer,fall,winter$]$, \review{with the simplifying
  assumption that spring and fall are equivalent $(\beta_1 =
  \beta_3)$}. Temperature data from the Swedish Meteorological and
Hydrological Institute was used to locally determine the duration of
the seasons.

\subsection{Data-driven simulations and pathogen data}
\label{subsec:data}

While there are several options to emulate network contact details, we
have utilized explicit data from the Swedish national cattle database
\cite{noremark2011network, siminf1}. The database contains information
about the individual animals in the population, including birth/death
dates and movement records. The data were transformed into anonymized
events for eight years of simulation over \numprint{37220} nodes, and
consist of \numprint{5470039} events, of which \numprint{624493}
events concern movements (\figref{fig:modelIntro}~(b)) and
\numprint{4845546} events are demographic. This type of data is
commonly recorded for livestock populations in many countries
\cite{BrooksPollock2014}, and is here relied upon to connect the nodes
by a transport network such that the overall spread mechanism is of
\emph{hybrid} character \cite{zhang2015optimizing}.

Without limiting the generality of the discussion, epidemiological
measurements can be regarded as a filter operating on the full
epidemiological state. In our case this filter simply returns a binary
answer 0 or 1 \emph{per node}: infection not detected/detected, and we
assume that a probabilistic model for this response is available and
can be simulated \textit{in silico} as a black box. The available
observations of STEC O157 \review{were collected} from a subset of 126
nodes on a bi-monthly basis during the time frame 2009--2012
\cite{vtec125}, \review{and the results were aggregated per quarter
  year, see \figref{fig:modelIntro}~(c)}. Each observation included
several bacteriological samples from the farm environment resulting in
a negative or positive detection per node. To truthfully mimic this
\textit{in silico} we rely on an urn model driven by empirically known
sensitivities given an underlying node prevalence. Details concerning
the pathogen detection protocol and the associated urn model are found
in the Supporting Information (SI).

\subsection{Approximate Bayesian Computations}
\label{subsec:ABC}

The simulation-driven Bayesian setup can be summarized as follows. We
regard the epidemiological model as a stochastic process
$X_t = X_t(\theta)$ \review{in continuous time}, dependent upon some
parameter $\theta$\review{$ = [\theta_1,\theta_2,\ldots]$}. Data from
this process is collected via some measurement filter $M(\cdot)$
operating on the state of the process \review{at time $t_i$}, i.e.,
$x_i = M(X(t_i;\;\theta))$, and where the filter itself might be an
additional source of noise. We assume that there is a simulator
$F(\cdot)$ that simulates \emph{and} measures the process such that
data may be sampled \textit{in silico} provided that a parameter value
$\theta$ is proposed, e.g., $z \sim F(\theta)$ simulates new data
$z = (z_i)$. The likelihood $\Pr(\cdot|\theta)$ is generally
computationally intractable\review{, although in principle the problem
  can be formulated as a missing data problem (see
  e.g.~\cite{lau2017mechanistic}), these are typically very
  computationally intensive and difficult to implement. Instead, an
  alternative approach is to use simulation-driven approximate
  Bayesian methods which find an approximate posterior distribution.}

The basic ABC rejection sampler \cite{beaumont2002approximate} accepts
proposed parameters depending on the output from a kernel function
$K_\varepsilon(\|z - x\|)$ with distance $\varepsilon \ge 0$. A common
approximation step is the use of \emph{summary statistics} $s = S(z)$,
effectively reducing the dimensionality of the data to compare;
instead of measuring the distance between the full data, one uses the
summary statistics of the data. The probability that we accept a
parameter proposal from a prior $\theta' \sim P(\theta)$ is then given
by $K_\varepsilon(|| S(z) - S(x)||)$ where $z \sim F(\theta')$. The
problem of choosing suitable summary statistics is discussed in
\cite[Chap.~5]{sisson2018handbook}. We select our summary statistics
similarly to previous suggestions for systems of comparable seasonal
characteristics \cite{papamakarios2016fast,
  everitt2018bootstrapped}. The statistics we use for our data,
\review{a time series of the number of aggregated positive samples per
  quarter year,} are the mean prevalence per quarter and the two
largest in magnitude Fourier coefficients, all in all 6 summarizing
statistics coefficients.

For observed summary statistics $s_{\text{obs}} = S(x)$, the resulting
ABC posterior, $\Pr_{\text{ABC}}(\theta|s_{\text{obs}})$, has the form
\begin{equation}
  \Pr_{\text{ABC}}(\theta | s_{\text{obs}}) \propto
  \int K_\varepsilon(||s -  s_{\text{obs}}||)
  \Pr(s| \theta) P(\theta) \, ds,
\end{equation}
where $\Pr(s| \theta)$ is the likelihood of $s = S(z)$ implied by
$\Pr(z|\theta)$. If $\varepsilon \rightarrow 0$, then by the
properties of the kernel function,
$ \lim_{\varepsilon\to 0}{\Pr_{\text{ABC}}(\theta|s_{\text{obs}})}
\propto \Pr(s_{\text{obs}}|\theta) P(\theta)$. Hence the posterior is
exact provided the summary statistics are sufficient, i.e., that no
information is lost \cite[Chap.~1]{sisson2018handbook}. However, this
is commonly not the case and the ABC posterior is then only
approximate.

\subsection{Adaptive Markov chain Monte Carlo and Synthetic likelihoods}
\label{subsec:SLAM}

A downside of ABC rejection is its comparably low acceptance
rate. Successful attempts to address this have been achieved by
combining the ABC methodology with existing likelihood-based sampling
methods, e.g., Metropolis-Hastings \cite{marjoram2003markov}. In the
seminal work \cite{wood2010statistical}, it was observed that measured
summary statistics are often asymptotically normally distributed,
i.e., $s \sim \mathcal{N}(\mu, \Sigma)$ with mean $\mu$ and covariance
$\Sigma$. The (limiting) likelihood of the observed summary statistic
is then the probability $\mathcal{N}(s_{\text{obs}}| \mu, \Sigma)$,
usually referred to as the synthetic likelihood (SL). Since $\mu$ and
$\Sigma$ are not known they must be estimated, and a natural idea is
to replace them by sample estimates from multiple simulations of $z$
with the same proposal $\theta^*$:
\begin{align}
  \label{eq:sl}
  \mathbf{S} &= (S_i) = (S(z_1), S(z_2), \dots, S(z_N)), \qquad z_j  \sim F(\theta^*), \\
  \hat{\mu} &= \frac{1}{N} \sum_{i=1}^N S_i, \label{eq:sampleMean}\\
  \hat{\Sigma} &= \frac{1}{N-1} \sum_{i=1}^N(S_i - \hat{\mu}) {(S_i -
                 \hat{\mu})}^\top,
                 \label{eq:sampleCovariance}
\end{align}
for which the log-SL is
\begin{equation}
  \label{eq:SL}
  \log \mathcal{N}(s_{\text{obs}} | \hat{\mu}, \widehat{\Sigma}) =
  -\frac{1}{2}{(s_{\text{obs}} - \hat{\mu})}^\top
  \widehat{\Sigma}^{-1}(s_{\text{obs}} - \hat{\mu}) -
  \frac{1}{2}\log|\widehat{\Sigma}|-\frac{N \mbox{dim}(s)}{2} \log(2\pi).
\end{equation}

Now consider the SL as $N \rightarrow \infty$. Under broad assumptions
\eqref{eq:sampleMean}--\eqref{eq:sampleCovariance} will converge to
$(\mu,\Sigma)$. However, the limit can be computationally impractical
for models which are expensive to simulate. An approach to improve the
estimate of SL is by using the bootstrap \cite{efron1979}, and
\cite{everitt2018bootstrapped} recently proposed an empirical
bootstrapping procedure designed specifically for SL-driven
algorithms. We found that it successfully produced more robust
estimates of the SL using fewer model calls.  The idea is to first
determine an empirical distribution $\hat{F}_N(\theta^*)$ by using $N$
independent simulations as follows. We compute synthetic data
$(z_{ij})$, where $i = 1,\ldots,N_{\text{time}}$ runs over time as
before and $j = 1,\ldots,N$ runs over the number of independent
trajectories used. Notably, the empirical distribution is a
distribution over $\Realdom^{N_{\text{time}}}$ and is constructed by
assuming each trajectory to yield $N_{\text{time}}$ \emph{independent}
samples, one per each point in time. \review{While this assumption is
  not exactly fulfilled, the measurements are well separated in time
  such that independence should be at least an accurate
  approximation}.

From the empirical distributions, we next sample $R \gg N$ new such
time series by \review{iterating through the $N_{\text{time}}$ time
  points, independently sampling with replacement from each of the $N$
  recorded simulated trajectories, and building up in this way each
  sampled time series}. The use of the bootstrap scheme thus results
in a larger dataset, effectively bootstrapping the estimate closer to
the desired limit of $N \rightarrow \infty$. Practically, the sample
sizes used in our experiments were $N = 20$ and $R = 100$.

The SL can be used in any desired likelihood-based inference
method. We chose to implement the SL in an Adaptive Metropolis (AM)
scheme \cite{haario2001adaptive}. Instead of sampling proposals from a
fixed prior distribution, the AM scheme samples adaptively using a
Gaussian with covariance \review{matrix} computed from the previous
entries in the Markov chain \cite{haario2001adaptive},
\begin{equation}
  \label{eq:recCov}
  C_{i+1} = \frac{i-1}{i}C_i + \frac{\xi_d}{i} \left(
  i \bar{\theta}_{i-1} \bar{\theta}^{\text{T}}_{i-1} -
  (i+1)\bar{\theta}_i \bar{\theta}^{\text{T}}_i +
  \theta_i \theta^{\text{T}}_i +
  \epsilon I_d \right),
\end{equation}
which does not add significant computational time, since the running
mean of the chain $\bar{\theta}_{i}$ can be computed recursively. In
\eqref{eq:recCov}, $\xi_d$ is a tuning parameter for the proposal
distance, as in regular Metropolis \cite{metropolis1953equation}, and
one uses a small value $\epsilon$ \review{to prevent a possible
  degeneration} of $C_i$. The adaptivity results in the loss of the
Markov property between samples, however, the chain keeps the desired
ergodic properties of its non-adaptive counterpart
\cite{haario2001adaptive, andrieu2006ergodicity}.

We refer to the combined algorithm as Bootstrapped Synthetic
Likelihood Adaptive Metropolis (SLAM) (\algref{alg:SLAM}).

\begin{algorithm} {
    \caption{\textit{Synthetic Likelihood Adaptive Metropolis (SLAM).}}
    \label{alg:SLAM}
    \begin{algorithmic}[1]
      \Require{Summarized data $s_{\text{obs}}$ and initial guess
        $(\theta_1,\mathcal{L}_{\theta})$. Adaptivity parameters
        $i_0$, $C_0$, $\xi_d$, and $\epsilon$ \review{(see text for
          details)}.}
      \For{$i = 2, \dots, N_{\text{sample}}$}
      \State{\textbf{if} $i > i_0$ \textbf{then}
        Compute $C_i$ by \eqref{eq:recCov}
        \textbf{else} $C_i = C_0$}
      \State{Sample
        $\theta^* \sim \mathcal{N}(\theta_{i-1},C_i)$}
      \State{Simulate
        $Y = \big( y_1, \dots, y_N \big), y_j \sim F(\theta^{*})$}
      \State{Bootstrap
        $Z = \big( z_1, \dots, z_R \big), z_j \sim \hat{F}_N(Y)$}
      \State{Estimate
        $(\hat{\mu}_{\theta^*}, \widehat{\Sigma}_{\theta^*})$ from
        $S = \mathbf{S}(Z)$, with \eqref{eq:sampleMean} and \eqref{eq:sampleCovariance}}
      \State{Compute
        $\mathcal{L}_{\theta^*} = \Pr_{\theta^*}(s_{\text{obs}}|S)$ by \eqref{eq:SL}}

      \If{$\text{Uniform}(0,1) < \min \big( 1,\mathcal{L}_{\theta^*} /
        \mathcal{L}_\theta\big)$}
      \State{$\theta_i = \theta^*$ and
        $\mathcal{L}_\theta = \mathcal{L}_{\theta^{*}}$}
      \Else
      \State{$\theta_i = \theta_{i-1}$}
      \EndIf
      \EndFor{}
    \end{algorithmic}}
\end{algorithm}

After running the adaptive Metropolis sampler for $N_{\text{train}}$
proposals using $P$ independent parallel replicas\review{, all with
  independent initial starting points}, the AM has explored the
support of the posterior. To refine the sample resolution, we next
deploy a Metropolized Independent Sampler (MIS)
\cite{hastings1970monte}, for $N_{\text{sample}}$ proposals, again
running in $P$ independent parallel replicas. These replicas all use a
single static normal proposal density, \review{namely the aggregate of
  the training replicas (mean of normal means and covariances after
  removal of a burn-in transient).} \review{The use of a MIS lowers
  the correlation between samples and shortens the burn-in
  time. Additionaly, using a prior adapted proposal distribution, the
  MIS sample rate is more resource efficient than the training phase.}
This scheme of combining AM and MIS is remindful of the block-adaptive
strategy proposed in \cite{jacob2011using}, \review{however, without
  any further adaptation after the first block, see \algref{alg:SLAM+}
  in the SI.}

\subsection{Parametric bootstrap and estimator efficiency}
\label{subsec:error}

The target of a Bayesian approach to parameterization is the parameter
posterior distribution $\Pr^*$ and its samples $\Theta^* \sim \Pr^*$.
In practice, the best one can hope for is approximate posterior
samples $\tilde{\Theta} \sim \tilde{\Pr}$, obtained from a numerical
model which approximates the processes underlying the data, and using
some approximate posterior sampling procedure. The error in the
Bayesian posterior estimator is then formally the difference
$\tilde{E} := \tilde{\Theta}-\Theta^*$, but without any useful
dependency relation between $\Pr^*$ and $\tilde{\Pr}$, one in practice
seeks to quantify some statistics of this error. Useful such measures
are typically derived from a point estimator of $\Theta^*$ and we
consider the minimum mean square error estimator (MMSE) for this
purpose, which is just the mean of the true posterior, $\theta^* :=
\Expect[\Theta^*]$. The mean square error is then
\begin{align}
  \label{eq:MSEdecompose}
  \tilde{e}^2 &:= \Expect[(\tilde{\Theta}-\theta^*)^2] =
      \underbrace{\Expect[(\tilde{\Theta}-\tilde{\theta})^2]}_{\text{Variance}}
              +\underbrace{(\tilde{\theta}-\theta^*)^2}_{\text{Square bias}},
\end{align}
where $\tilde{\theta} := \Expect[\tilde{\Theta}]$ is the MMSE of
$\tilde{\Theta}$. This decomposes the mean square error into the
variance of $\tilde{\Theta}$ and the square of the \emph{bias}
$\tilde{b} := \Expect[\tilde{E}] = \tilde{\theta}-\theta^*$.

The procedure we propose for estimating the bias falls under the class
of methods referred to as \emph{parametric bootstrap}
\cite{efron1979}. The general idea is to treat inference
about $\Pr^*$ for the original data as comparable to inference of
$\tilde{\Pr}$ for resampled synthetic data. We thus use the same
Bayesian posterior sampling used to sample from $\tilde{\Pr}$ a second
time and produce samples
$\tilde{\tilde{\theta}} \sim \tilde{\tilde{\Pr}}$, that is, from the
posterior given synthetic data generated from known parameters. The
synthetic data may be generated by either parameters drawn from
$\tilde{\Pr}$, or from an associated point estimator. We simply used
the MMSE for this step and generated data using the sample mean from
the posterior $\tilde{\Pr}$, that is, for
$\hat{\theta} := N_{\text{sample}}^{-1} \sum_i \tilde{\theta}_i$,
where $(\tilde{\theta}_i)$ are samples from $\tilde{\Pr}$. The error
in this data-synthetic inference is
$\tilde{\tilde{E}} := \tilde{\tilde{\Theta}}-\hat{\theta}$, and hence
its bias is $\tilde{\tilde{b}} := \Expect[\tilde{\tilde{E}}]$, readily
estimated as a sample mean. The bootstrap estimate of the wanted bias
is then simply
$\tilde{b} \approx \tilde{b}_{\text{BS}} := \tilde{\tilde{b}}$, which
together with the sample variance of $(\tilde{\theta}_i)$ yields the
bootstrap estimate of the mean square error as in
\eqref{eq:MSEdecompose}.

For our data we generated $M_{\text{boot}} = 10$ posterior
distributions $\tilde{\tilde{\Pr}}$ using independent model data
realizations from the same parameter $\hat{\theta}$.  Each
distribution consisted of $N_{\text{sample}} = 45$ samples, after
removal of burn-in ($= 500$) and thinning (every $100$th). Finally,
the estimated bias was computed as the average bias across the
$M_{\text{boot}}$ bootstrap posteriors.


\section{Results}

We first look into the issue of the model's identifiability by working
through a set of synthetic set-ups on smaller scale using known-truth
data. The rationale here is that, if this phase does not succeed,
there will be little hope in coping with more realistic
situations. Results on this are reported in
\S\ref{subsec:identifiability} where, as a side-effect, we also
quantify the efficiency of the SLAM compared to the basic ABC
rejection procedure. Our main results are found in
\S\ref{subsec:national} where a full national scale STEC O157 model
based on first principles is parameterized from available pathogen
data. Applications concerning detection and intervention scenarios are
exemplified in \S\S\ref{subsec:detection}--\ref{subsec:intervention}.

\subsection{Identifiability and the efficiency of SLAM}
\label{subsec:identifiability}

Given real-world pathogen data and a disease spread model with an
intractable likelihood, the options to rigorously analyze the
\review{parameters'} identifiability are quite limited. \review{A
  practical approach is to evaluate how well a proposed inference
  method performs on synthetic data drawn from a known model.} By
testing different inference procedures on a problem with an
established truth \review{and comparing their} estimation qualities, we
empirically reveal both identifiability and estimator efficiency. In
the Bayesian setting we may, for example, compare the error in some
point estimator derived from the posterior, e.g., the average of the
posterior samples (the minimum mean square error estimator). If the
stability of the inference procedure remains when confronted with real
data, then we may use the bootstrap procedure discussed in
\S\ref{subsec:error} \review{to more} rigorously assess the quality of the
posterior.

Our full scale national model is computationally expensive,
particularly so considering that parameterization requires large
quantities of sample trajectories. To more effectively explore the
limits of the methodology, we initially consider a synthetic set-up at
a smaller scale. We define this scaled down model over \numprint{1600}
nodes, using synthetically generated movements of individuals for four
years while recording the full epidemiological state every 60th day on
a sample of 100 randomly chosen nodes \review{(see the SI for details
  on how this was done in practice)}. We now ask if this synthetic
data can be used to reliably infer the parameters used in the local
infection model: $\upsilon$, $\beta$, and $\gamma$. We artificially
construct some seasonality by using two values of $\beta$, that is,
$\beta_1$ and $\beta_2$ each cover a 6-month period of the year, and
we consider this modeling choice to be part of the prior information
together with precise knowledge of the initial state at time $t = 0$.
This synthetic set-up is fast to simulate but preserves many of the
main characteristics of the full national scale model.

We supply this system with parameters which, after some initial trial
and error, were found to be reasonably close to the domain of
relevance for the STEC O157 application later considered. Data from a
single simulation was fed into an off-the-shelf ABC rejection sampler
as well as into our own SLAM posterior sampler. The procedures were
initially given ``unfiltered'' data in the form of the exact disease
prevalence, i.e.~the fraction of infected individuals, at the 100
selected nodes and at the sample points in time. Later, we moved on to
\emph{binary filtered} data obtained by subjecting the full state to a
computational model of a certain pathogen detection protocol, see the
discussion in \S\ref{subsec:data} with further details in the SI
(cf.~also \figref{fig:modelIntro}~(c)). Intuitively, one expects a
hopefully acceptable loss of estimator accuracy when switching to
filtered data.

\begin{figure}[htp]
  \centering
  \includegraphics[width=1\linewidth]{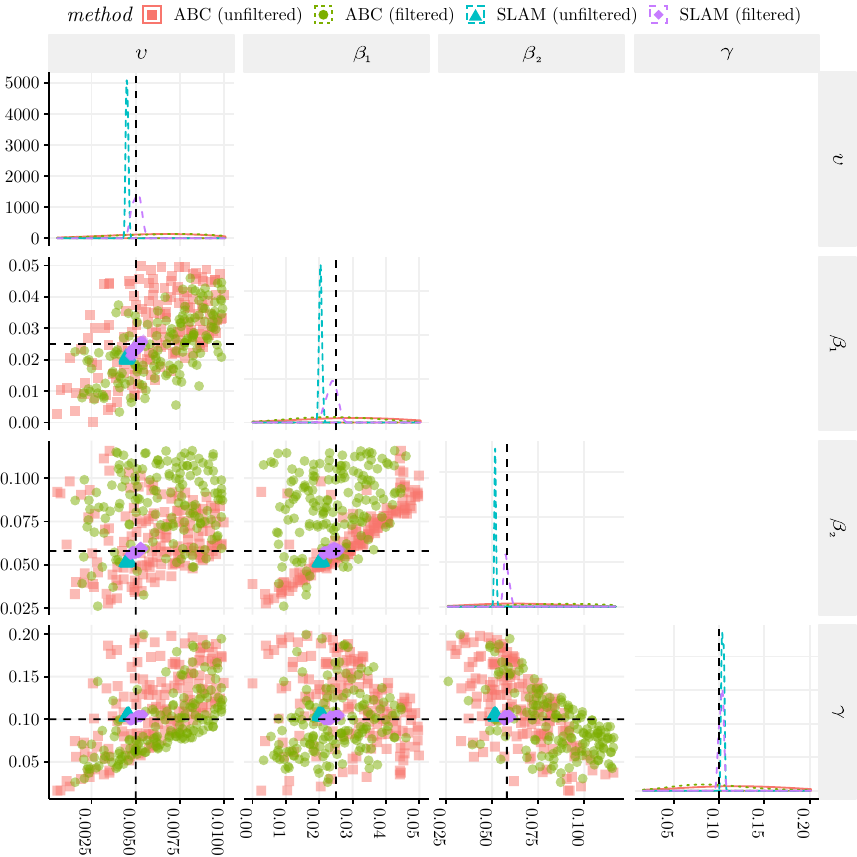}
  \caption{Posterior samples for the \numprint{1600} nodes synthetic
    dataset (\review{positive uniform unnormalized} priors). The
    samples\review{, indicated by distinct colors and shapes,} are
    generated by either ABC rejection (\review{on} unfiltered \&
    binary filtered data) or SLAM (\review{on} unfiltered \& binary
    filtered \review{data}). The true parameters are indicated by the
    \review{horizontal and vertical black \& dashed lines}. For
    comparable results both methods were run the same amount of
    wall-clock time. \review{The plots on the diagonal are the
      marginal posterior distribution per parameter. The lower
      diagonal plots are bi-dimensional posterior samples. The axis
      show the parameter values, except for the top left plot's
      $y$-axis which is the density.}}
  \label{fig:1600MultiSmall}
\end{figure}

The resulting posteriors are shown in \figref{fig:1600MultiSmall}. All
samplers were successful in this synthetic setting. For SLAM,
filtering the data implies a small increase in posterior width,
\review{and a minor improvement of the mean posterior bias. Although
  the effect in bias is a bit surprising, the difference is quite
  small and, moreover, a shift in bias is suitable to handle through
  bootstrapping estimates as is done in \S\ref{subsec:national}}. The
quality of the ABC sampler is somewhat more difficult to assess. Since
unfiltered data is a best possible scenario, this case also defines an
upper bound of the posterior quality. For the ABC sampler, the
filtered state observations yield a posterior which is seemingly
unaffected by the data being filtered. Since proposals are accepted
under an $\varepsilon$-criterion, and since with different kinds of
data, the meaning of $\varepsilon$ varies, \review{the unfiltered and
  filtered versions} can not be directly compared. Perhaps more
interestingly, the error of the SLAM MMSE estimator is almost one
order of magnitude smaller than that of the corresponding
ABC-estimator (see the SI for exact errors), conditioned on using the
same amount of wall-clock time. These results show that the transition
from a straightforward ABC rejection sampler to a posterior obtained
via a synthetic likelihood ansatz \review{can pay off well}.

\subsection{National scale Bayesian inversion}
\label{subsec:national}

As we now show, SLAM is successful in inverting a full national scale
model of the spread of STEC O157 using the pathogen data from
\cite{vtec125}. We consider anew the SIS$_E$ model in
\figref{fig:modelIntro}~(a), now replicated across \numprint{37220}
actual nodes (herds) populated by about 1.6 million individual
animals, and connected by the full eight year national scale animal
movement dataset over Sweden (\figref{fig:modelIntro}~(b)). The model
parameters are $(\upsilon, \beta_1, \beta_2, \beta_4, \gamma, p_0)$,
where $(\beta_{1,2,3,4})$ is the decay of infectious matter for
$[$spring,summer,fall,winter$]$, and $\beta_1 = \beta_3$ to
simplify. As mentioned, \review{seasons were inferred} on a per-node
basis using climate data from the Swedish Meteorological and
Hydrological Institute. The parameter $p_0$ is a compact description
of the system's initial state as follows. On initialization the total
number of individuals $S_0+I_0$ at each node is known from data, and
we first sample $I_0$ to match any proposed prevalence $p_0$:
$I_0/ (S_0+I_0) \approxeq p_0$. Effectively assuming equilibrium
($\varphi'(0) \approx 0$ in \eqref{eq:_E}), we next set the
\review{node's initial environmental infectious pressure} to be
$\varphi_0 = \beta^{-1} I_0/ (S_0+I_0)$.

The pathogen data from \cite{vtec125} is understood as the result of
the binary filter applied to the full state of the tested node. The
measures are low-informative on the epidemiological state and
distributed sparsely in both space and time: about 6--8 binary
true/false samples per year, and only at 0.3\% of the nodes. The
computational complexity is a significant difficulty of any realistic
inverse problem. For each parameter proposal, we estimate the
synthetic likelihood from $N = 20$ trajectories, bootstrapped to $R =
100$. One trajectory results from about $10^8$ simulated events such
that, in the end, each proposal is evaluated in about 60s over 16
compute cores ($2 \times \text{Intel Xeon E5--2660}$). \review{Similar
  to the inversion of the \numprint{1600} node network in
  \S\ref{subsec:identifiability}, before using any real observations,
  we tested the feasability using synthetic known-truth observations.}

\begin{figure}[htp]
  \centering
  \includegraphics[width=1\linewidth]{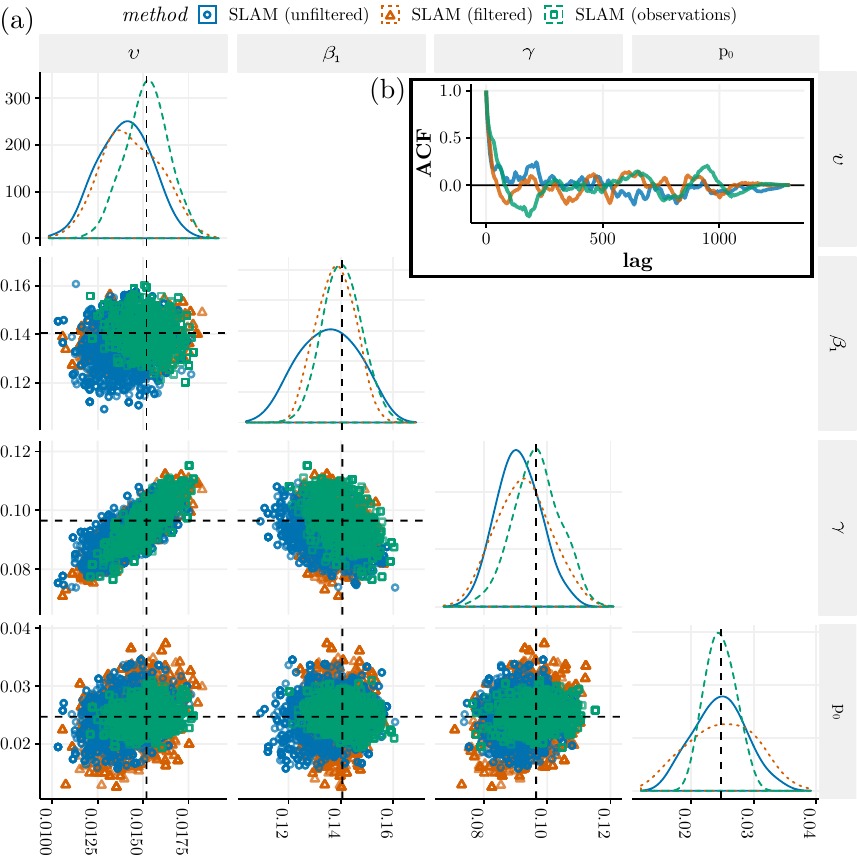}
  \caption{{\bf (a)} Posterior parameter samples obtained from SLAM
    given synthetic unfiltered data, binary filtered data, and real
    pathogen observations (\review{positive uniform unnormalized
      priors}). The multivariate Gelman and Rubin's convergence
    diagnostic for each method is [1.05, 1.09, 1.07]. A version of the
    figure including all $\beta$ dimensions is found in the
    SI. \review{The horizontal and vertical black \& dashed lines are
      the MMSE estimators from SLAM (observations): $(\upsilon,
      \beta_1, \gamma, p_0) = (0.0151, 0.141, 0.0965, 0.0247)$.} {\bf
      (b)} The autocorrelation of the samples from one of the parallel
    chains of the parameter $\upsilon$.}
  \label{fig:RealMultiFull}
\end{figure}

%
%
We obtain the approximate posterior parameter distribution of the STEC
O157 endemics in \figref{fig:RealMultiFull}. We next perform
parametric bootstrap as discussed and improve on our confidence in the
results by leveraging the stability of the inference procedure as
follows. The posterior mean is used as a suitable parameter to
replicate new synthetic data for, and the inference procedure is put
to work again using this data. Here we apply SLAM both to the
synthetic unfiltered data as well as to the filtered version. As seen
in \figref{fig:RealMultiFull} the posterior distributions all match
very closely. We quantify the accuracy of the minimum mean square
error (MMSE) parameter point estimator through the MSE for the three
posteriors. The bias is unknown for the most interesting case of real
observations and we there have to impute the bias via the
corresponding synthetic quantities, resulting in a
bootstrapped/imputed estimated accuracy of the point estimator
\cite{efron1979}. All MMSE parameter estimators were found to be
within 10\% accuracy and in most cases within 5\% (see SI).

\begin{figure}[htp]
  \centering
  \includegraphics[width=0.99\linewidth]{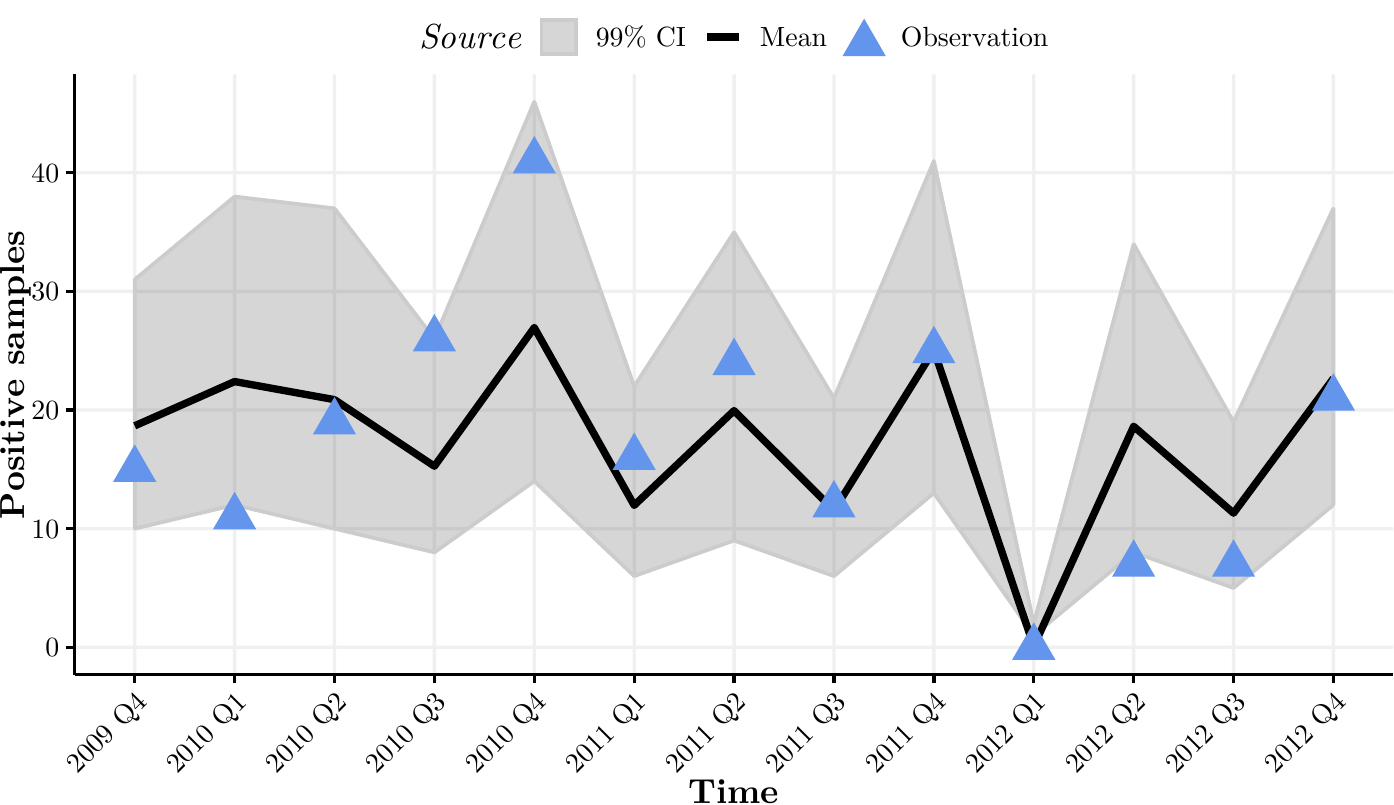}
  \caption{\review{Posterior predictive check:} the observations of
    the full system and \review{$N = 2000$} posterior sample
    trajectories. The blue triangles are the observations of the
    number of aggregated positive \review{samples per quarter}, the
    solid black line is the mean of the posterior \review{samples} and
    the shaded area is the \review{99\%} credible interval.}
  \label{fig:posteriorFit}
\end{figure}

The posterior fit to data can be judged by comparing our estimated
node prevalence of $11.2 [7.9, 15.5]\%$, at $95\%$ credible interval
(CI) from $N = 250$ posterior parameters, to the reported value
$13.1\%$ \cite{vtec125}. We further validated our posterior by looking
at two independent observational studies. We estimated the temporal
\review{population prevalence to} $2.2 [1.5, 3.5]\%$ for (2008--09)
and, \review{respectively, to} $2.2 [1.5, 3.3]\%$ (2011--12) ($N =
250$, $95\%$ CI). These measures compare well to the studies
\cite{vtec2008_09, vtec2011_12}, where those figures are $3.3\%$ and
$3.1\%$, respectively. In \figref{fig:posteriorFit} the posterior
simulator is also compared more directly to data.

For the parameters, our accuracy can be summarized as an average
coefficient of variation of around 7\%, while previous comparable
attempts obtain around 30\% \cite{ferguson2003planning,
  ferguson2005strategies, fournie2018dynamic, merler2011determinants,
  brouwer2018epidemiology, Brooks-Pollock2014ABCSMC}, albeit in most
cases for point estimators (a notable exception is
\cite{Brooks-Pollock2014ABCSMC}, \review{which also fit a stochastic
  national scale livestock disease model incorporating movement data
  within a Bayesian (ABC) framework}). From our experience, the most
decisive steps towards the quality of our results were (1) the good
topological agreement in the prior model formulation through the
combined effects of detailed network data and a local
climate-dependent $\beta$, (2) the effective parameterization of the
initial state through the scalar $p_0$, and (3), the use of empirical
bootstrapping to increase the robustness in estimating the synthetic
likelihood. Additionally, using adaptive Metropolis proposals
increased the sampling efficiency considerably.

\subsection{Detection: large nodes are nearly optimal sentinels}
\label{subsec:detection}

The number of human STEC O157 cases is less than about 10 per
\numprint{100000}, but the risk for children is considerably higher
than for adults. Infected individuals often develop bloody diarrhea,
and about 5--10\% further develop hemolytic-uremic syndrome, a severe
complication that can be fatal \cite{parry2000public}. By reviewing
surveillance routines and mitigation strategies, one could reduce the
prevalence of STEC O157 in reservoirs, e.g., in the cattle population,
thus potentially reducing the number of human infections
\cite{bajardi2012optimizing, siminf1, keeling2017efficient}. The
process of selecting sentinel nodes has been actively studied
\cite{keeling2005networks, bajardi2012optimizing,
  scharrer2015evaluation}, and here we propose a way by which the
sensitivity of the pathogen detection procedures can be improved
within our framework.

We test five different kinds of \emph{sentinel} node sets, each of the
same size (10 nodes), and evaluate their detection sensitivity through
multiple simulations, using $N = 250$ sample parameters from the
previously computed parameter posterior. The first four sets are
defined in terms of simple network measures, namely indegree
\review{(Indegree)}, outdegree \review{(Outdegree)}, node population
\review{(Largest)}, and, for reference, a random set
\review{(Random)}. We design the fifth set \review{(Observation)} to
be nearly optimal as follows. We first simulated independent
trajectories for eight years, while recording the infectious pressure
$\varphi$ during the final four years. By ranking the nodes in the
system according \review{to the environmental pathogen load}
$\varphi$, we obtained a shortlist of the \review{10} most infected
nodes, and we let these serve as a close to optimal detection set of
nodes.

For the evaluation of the node detectability, the system is simulated
using parameters from the posterior distribution. We define the
detection sensitivity for the sentinel nodes by using a simple
detection probability model as follows.  At node set $\mathcal{N}_i$
and at time $t$, compute
$\Pr(\text{detect infection}) = 1-\prod_{i \in \mathcal{N}_i}
(1-P_i(\varphi_i(t)))$, where $P_i(\cdot)$ is modeled by a sigmoid
function, $P_i(x) = 1/[1 + \exp(-k(x-\varphi_0))]$, where $k$ is the
test's sharpness and $\varphi_0$ the cut-off. Although our choice of
sigmoid parameters is not important for our purposes and were
arbitrary, the modeling choices here can clearly be made to mimic any
empirically known sensitivity.

\begin{figure}[htp]
  \centering
  \includegraphics[width=1\linewidth, angle = 0]{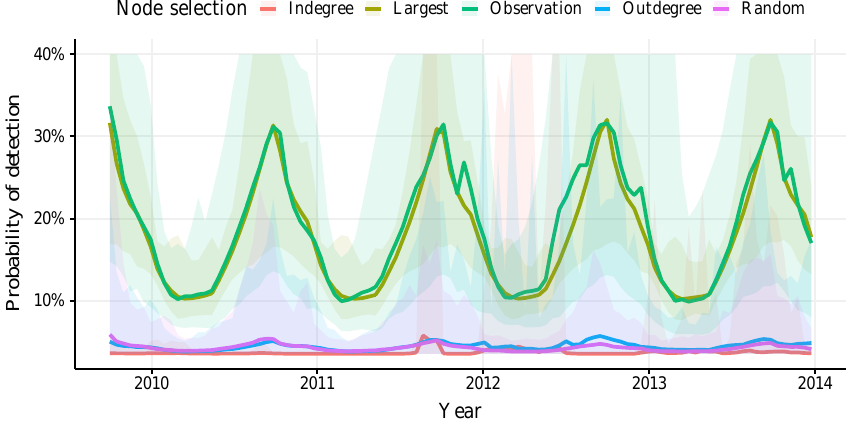}
  \caption{The detection probability over time for five different sets
    of sentinel nodes. The node sets \emph{Indegree}, \emph{Largest},
    and \emph{Outdegree} were defined from basic network statistics,
    and \emph{Random} using uniform node sampling. The
    \emph{Observation} set is found through pre-simulation and is
    nearly optimal. Each set included the 10 highest ranking nodes
    according to each criterion. Displayed is the mean detection
    probability and 95\% CI. \textit{Sigmoid model parameters:}
    $k = 15$ and $\varphi_0 = 0.375$, for which
    $\varphi \in [0.13,0.62]$ forms a symmetric 95\% CI.}
  \label{fig:detection}
\end{figure}

The measured detection probability, displayed in
\figref{fig:detection}, shows that the node set \emph{Largest} is
about as efficient as the upper limit estimate \emph{Observation}. An
explanation for the apparent efficiency of population size as a
priority measure for detection procedures lies in \emph{the rescue
  effect} \cite{keeling2008modeling}, which states that, in a
metapopulation with high interconnectivity, the larger nodes take the
role of ``rescuing'' the disease from extinction; an early observation
of this phenomenon was made in \cite{finkenstadt2002stochastic} in a
study of measles. The largest nodes tend to remain continuously
infected such that the infectious pressure is larger than
elsewhere. This is also consistent with large groups of cattle being
more likely to be STEC positive \cite{vidovic2006prevalence,
  ellis2007identification}. The marginal performance differences
between sets based on outdegree and indegree is in line with previous
findings \cite{bajardi2012optimizing}, that is, they do not show a
significant improvement to using randomly selected nodes. Note also
the clear periodic trend which could well be exploited to propose
further refined detection strategies.

\subsection{Intervention: best results from local actions}
\label{subsec:intervention}

An important purpose with a computational epidemiological framework is
to asses the effects of interventions. In
\cite{merler2015spatiotemporal, peak2017comparing} the efficiency of
mitigation strategies for epidemics were evaluated using data-driven
agent-based models, and a Bayesian node-based framework was employed
for the same purpose in \cite{Brooks-Pollock2014ABCSMC}. For STEC O157,
\cite{siminf3} proposes and assesses various interventions, although
not in a probabilistic framework. We test comparable intervention
proposals in our Bayesian \emph{in silico} model.

We sample 250 model parameters from the posterior and simulate for
four years to avoid any transient effects. Then we intervene according
to three strategies as follows. The first strategy is to remove
transmission by transport, i.e., if an infected animal is moved we
change the status to susceptible on arrival. The second strategy is to
increase the bacterial decay rate $\beta$ by 10\%, and the third is to
reduce the indirect transmission rate $\upsilon$ by 10\%. These
interventions do not invalidate the recorded transport events and
therefore do not interfere with any implicit causality. After the
selected intervention we record the population prevalence for five
more years.

\begin{figure}[htp]
  \centering
  \includegraphics[width=1\textwidth, angle=0]{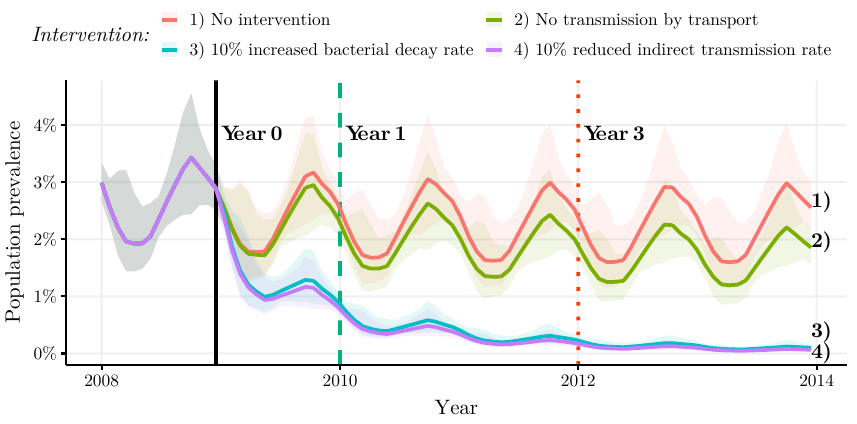}
  \caption{The prevalence response to intervention. Three different
    techniques were tested starting from the intervention point in
    time (solid vertical line). Displayed is the mean population
    prevalence and 95\% CI. Measured reduction rates for each
    intervention protocol are reported in the SI.}
  \label{fig:intervention}
\end{figure}

In line with the results of \cite{siminf3}, in
\figref{fig:intervention} we conclude that control strategies that
emphasize local intervention apparently give the best results, as
after three years the infection is controlled. Although the precise
interpretation of the local strategies needs to be defined, the
relative effects on the parameters could in principle be estimated
through experiments with practical procedures including, e.g.,
improved on-farm biosecurity or vaccination.

\section{Discussion}

Epidemiological models have both qualitative and quantitative
applications, and different trade-offs clearly apply depending on the
purpose of the model \cite{ferguson2003planning,
  keeling2008modeling}. In this paper we emphasized computational
models designed with the goal of assessing large-scale disease spread
in a quantitative sense. This is the regime of models for which data
is essential \cite{merler2015spatiotemporal, liu2018measurability,
  walters2018modelling}, both in driving the model, e.g., detailed
network data in our case, but also in parameterizing the model using
surveillance measurements. Given that pathogen surveillance data is
likely to become cheaper and more accessible, e.g., with improvements
in biosensors for livestock \cite{vidic2017advanced}, data-driven
models could facilitate timely and effective response to various
infections.

We have demonstrated the feasibility of a Bayesian parameterization on
a national scale using field data from the spread of STEC O157 in
Swedish cattle. The main modeling assumptions are that network data is
available and that the pathogen detection procedure can be replicated
\textit{in silico}. The use of a detailed transport pattern
facilitates model inversion thanks to the high level of topological
agreement already in the prior model formulation. For weakly
informative pathogen detection protocols, we developed ideas on
approaching the question of parameter identifiability via a series of
controlled synthetic data experiments. Our findings here support the
use of synthetic likelihoods and an adaptive Metropolis sampling over
the more straightforward ABC sampler. Once parameters have been
identified, parametric bootstrap enables an estimation of the
confidence in the inference procedure.

With the methodology developed here a substantial improvement in the
qualities of surveillance \review{and control} procedures in animal
and public health is possible. We have exemplified this in the ranking
of detection- and intervention procedures, where detailed credible
bounds are directly available thanks to the Bayesian framework. Our
work opens up \review{the potential} for better use of quantitative
large-scale epidemiological models whenever sufficient data is
available. With the increasing amounts of epidemiologically relevant
data being collected, we conclude that there is also an important role
to be played by models consistently driven by and informed from this
data.


\section*{Funding}

This work was financially supported by the Swedish Research Council
Formas (S.~Engblom, S.~Widgren), by the Swedish Research Council
within the UPMARC Linnaeus center of Excellence (S.~Engblom,
R.~Eriksson), and by the Swedish strategic research program eSSENCE
(S.~Widgren). The simulations were performed on resources provided by
the Swedish National Infrastructure for Computing (SNIC) at UPPMAX.

%

%

%

%

%

%



\appendix
\clearpage

\renewcommand{\Pr}{\mathbf{P}}

\newcommand{\Probspace}{\Omega}
\newcommand{\Probelem}{\omega}
\newcommand{\Probfiltr}{\mathcal{F}}

\newcommand{\X}{\mathbb{X}}
\newcommand{\Y}{\mathbb{Y}}

\newcommand{\Stoich}{\mathbb{S}}
\newcommand{\Connect}{\mathbb{C}}
\newcommand{\Stoichglobal}{\mathbb{G}}

\newcommand{\Nspecies}{M_{\mbox{{\tiny comp}}}}
\newcommand{\Ndet}{M_{\mbox{{\tiny conc}}}}
\newcommand{\Nnodes}{N_{\mbox{{\tiny nodes}}}}

\newcommand{\fatmu}{\boldsymbol{\mu}}
\newcommand{\fatnu}{\boldsymbol{\nu}}
\newcommand{\fatlambda}{\boldsymbol{\Lambda}}

\begin{center}
  \LARGE{\textbf{Supporting Information: \\ Bayesian epidemiological modeling \\
      over high-resolution network data}}
\end{center}

\renewcommand*{\citenumfont}[1]{SI#1}
\renewcommand*{\bibnumfmt}[1]{[SI#1]}

\let\oldcite\cite
\renewcommand*\cite[1]{\oldcite{SI-#1}}


The supporting information text consists of three parts. In the first
part, we discuss the details of the available epidemiological data in
the form of recorded population demographics and transport events, as
well as the STEC O157 measurement procedure that was used to produce
our pathogen data. The second part summarizes the ideas leading up to
our specific epidemiological model and the underlying simulation
engine. In the final part, we detail the Bayesian computational
protocols developed and highlight some additional supporting results.

\medskip

All associated data, computer codes, and experimental scripts are
available online at
\url{https://github.com/robineriksson/BayesianDataDrivenModeling}. The
\SimInf\ epidemiological framework is available at
\url{https://www.siminf.org}.

\section{Epidemiological data}

The data used for the full scale STEC O157 application study in the
paper consists of two datasets. Firstly, anonymized network events at
the level of single individuals, with \numprint{5470039} entries
covering enter-, external transfer-, and exit events
\cite{noremark2011network}. Secondly, anonymized infectious measurements of the
pathogen from a longitudinal observational study \cite{vtec125}, in
which 126 cattle holdings (nodes) out of a total of \numprint{37220}
were observed for 38 months in the period 2009--2013.

\subsection*{Network data}

We used the network data initially aggregated in
\cite{siminf1}\cite{siminf3}, which contains processed records from
the Swedish cattle database during the period in time 2005-07-01 to
2013-12-31 (102 months). The total number of cattle holdings was
\numprint{37220} which make up the epidemiological network of our
simulation. The network data falls into three types of events: enter,
external transfer, and exit. The enter events include births and
imports from abroad. The external transfer events are movements of
single individuals between the nodes of the network. The exit events
imply either slaughter, euthanasia, or export of the animal to another
country, i.e., the animal is removed from the observed network.

We followed an earlier prescription of handling the data in the
simulator \cite{siminf2}\cite{siminf1}. We randomly sample the
individuals from the compartments affected by the event without
consideration of the infection state and after moving the animals keep
the same disease state in the new holding as in the previous
holding. An assumption that was consistently made was that the
individuals who enter the network, i.e., are born or are imported, are
in the susceptible state.

\subsection*{Pathogen data}

As the data used for the parameterization of the national scale STEC
O157 model we relied upon the longitudinal observational study in
\cite{vtec125}. The study consisted of repeatedly collecting
environmental samples from 126 Swedish cattle holdings during the
period October 2009 to December 2012, with the aim of determining the
presence of STEC O157 at each herd and at several points in time.

\begin{SCfigure}
  \centering
  \includegraphics[angle=0,width=0.45\textwidth]{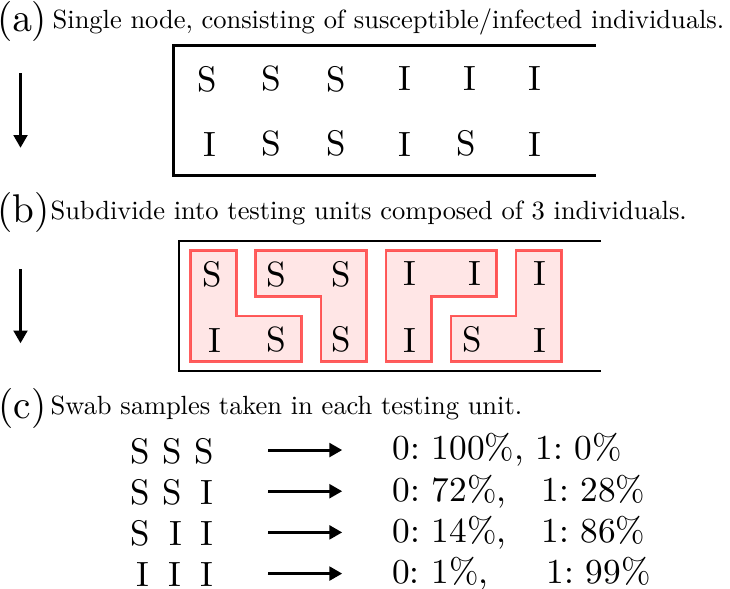}
  \caption{{\bf (a)} The population in a single node, consisting of
    susceptible (S) and infected (I) individuals. {\bf (b)} From the
    node population, testing units composed of 3 individuals are
    formed. {\bf (c)} Swab samples in each testing unit reveal
    infected individuals with empirically known sensitivities, 0: unit
    determined clean, 1: unit determined infected. The whole node is
    judged clean (0) provided none of the testing units are judged
    infected, and infected (1) otherwise.}
  \label{fig:urnmodel}
\end{SCfigure}

To determine the sample outcome of a simulated node, we replicated
\textit{in silico} the sampling protocol described in
\cite{vtec125}\cite{siminf1}. We proceed as follows. At the time of
the measurement and at the measurement site, we randomly form
\emph{testing units} consisting of three individuals sampled from the
pool of susceptible and infected individuals. The testing unit is
classified as clean or infected with an empirically known probability
\cite{cray1995}\cite{Widgren2013}, see \figref{fig:urnmodel}~(c). If
any testing unit is found to be infected, then the whole node is
classified as infected, otherwise the node is judged clean. The
protocol is schematically summarized in \figref{fig:urnmodel}.

\subsection*{Synthetic data}

The data for the scaled down model was generated to mimic the
seasonality in births, deaths and movements in the real livestock
data. For the animal movements, random connections between holdings
were used to generate the transport events. The synthetic data is
available in the \SimInf\ software, contains \numprint{466692} events
(exit = \numprint{182535}; enter = \numprint{182685}; external
transfer = \numprint{101472}) for \numprint{1600} holdings distributed
over $4 \times 365$ days.


\section{Epidemiological modeling}

This section develops the epidemiological modeling underlying the
results of the paper. The material is aimed to be as brief as possible
while still being self-contained. A less concentrated exposition can
be found in \cite{siminf_Ch}, see also \cite{siminf2} for some more on
the high-performance computing aspects. For general monographs,
consult
\cite{AnderssonBrittonSIR}\cite{BrauerEpidemiology}\cite{DiekmannBrittonMathEpidemics}\cite{KeelingModeling}. The
software \SimInf\ is detailed in
\cite{SimInf_software}\cite{siminf_manual} and is publicly available
at \url{www.siminf.org}.

\subsection*{Continuous-time Markov chains}

At the scale of a single node consisting of relatively few
individuals, say less than a few 100s, it is generally agreed that
stochastic effects can be important. A stochastic model may capture
the intrinsic noise due to the parts of the dynamics not explicitly
modeled, such as precise contact details of infectious agents. For a
discrete state variable, namely the number of individuals in a certain
epidemiological state, and in continuous time, this implies a
continuous-time Markov chain as the modeling framework of choice.

Consider first the classical SIR model \cite{Kermack1927},
\begin{align}
  &\left. \begin{array}{rl}
    S+I &\xrightarrow{\upsilon} 2I \\
    I &\xrightarrow{\gamma} R \\
  \end{array} \right\},
\end{align}
in terms of the integer state vector $X(t) = [S,I,R] =
[$\#Susceptible, \#Infected, \#Recovered$]$ individuals at time
$t$. The dynamics consists of two transitions: the first increases $I$
by one and decreases $S$ analogously, and the second similarly
decreases $I$ by one and increases $R$ by one. The transition rates
for these transitions are understood to be the associated
combinatorial product scaled with the corresponding rate parameter;
this is $\upsilon S I = \upsilon X_1(t)X_2(t)$ for the first, and
$\gamma I = \gamma X_2(t)$ for the second transition.

We may compactly write the stochastic SIR-model in the form of a
\emph{mass action} jump stochastic differential equation (jump SDE) as
follows. Assume a probability triplet $(\Probspace,\Probfiltr,\Pr)$
supporting Poisson processes for the different transitions. The
\emph{state vector} $X(t) = X(t; \, \omega) \in \Intdom_{+}^{3}$,
$\Probelem \in \Probspace$, counts at time $t$ the number of
individuals in each of the 3 compartments $S$, $I$, and $R$. Define
the \emph{stoichiometric coefficients} and the \emph{transition rates}
following the previously described logic of the SIR model,
\begin{align}
  \Stoich &= \left[ \begin{array}{rr}
      -1 & 0 \\
      1  & -1 \\
      0 & 1
    \end{array} \right],
  \quad
  R(x) = [\beta x_{1}x_{2},\gamma x_{3}]^{T}.
\end{align}
The Markovian steps are the columns of $\Stoich$ and their
corresponding intensities are the elements in $R$. A compact form for
the resulting mass-action continuous-time Markov chain is now
\begin{align}
  \label{eq:JSDE}
  dX(t) &= \Stoich\fatmu(dt),
\end{align}
where by $\fatmu(\cdot)$ we mean a state-dependent vector random
counting measure of deterministic intensity $R(\cdot)$,
\begin{align}
  \Expect[\fatmu(dt)] &= \Expect[\fatmu(x,dt)] = R(x) dt = [\upsilon x_1
    x_2,\gamma x_2]^T dt.
\end{align}
Epidemiological events are thus prescribed to arrive according to
competing Poisson processes of the corresponding intensities. An event
$r$ at time $t$ implies that the state is to be changed according to
the prescription $X(t) = X(t-)+\Stoich(:,r)$.

It is convenient to capture also events scheduled ahead of the
simulation with the same notation. The most important such event is
the physical transport of individuals between nodes, but also
demographic events can be modeled equivalently. \textit{A priori}
scheduled events at times $(t_{i})$ are thus associated with a sum of
temporal Dirac measures,
\begin{align}
  \mu_{3}(dt) &= \sum_{i} \delta(t_{i};\,dt),
\end{align}
together with an appropriate additional third column in $\Stoich$.

\subsection*{Mixed discrete-continuous states}

At a sufficiently large modeling scale there will usually be
concentration type variables for which an ODE-based description is
more natural. The obvious example is the concentration of bacteria in
an environment for which counting is impossible. A general mixed jump
SDE-ODE model ansatz is
\begin{align}
  \label{eq:JSDE_ODE}
  \begin{array}{rcl}
    dX(t) &=& \Stoich\fatmu(dt) \\
    Y'(t) &=& f(X(t-),Y(t)) \\
  \end{array}
\end{align}
where $(X,Y) \in \Intdom^{\Nspecies} \times \Realdom^{\Ndet}$ and
where the counting measure may now depend also on the concentration
variable; $\Expect[\fatmu(dt)] = R(x,y)dt$.

\begin{SCfigure}
  \centering
  \includegraphics[angle=0,width=0.35\textwidth]{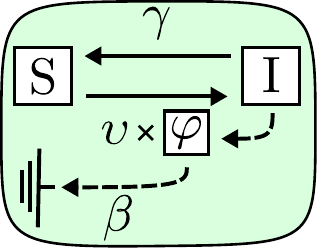}
  \caption{The SIS$_E$ three parameter model. Susceptible individuals
    $\text{S}$ turn infected at rate $\upsilon \times
    \varphi$. Infected individuals $\text{I}$ recover at a rate
    $\gamma$, and shed the pathogen to the environment adding to the
    infectious pressure $\varphi$, which in turn decays at rate
    $\beta$.}
  \label{fig:modelSchematic}
\end{SCfigure}

The model favored in the paper is the SIS$_E$ model which can be
defined from two integer states $X = [S,I]$, and one concentration
variable, $Y = \varphi$, the concentration of infectious
substance. The SIS$_E$ model follows from the specific choices
\begin{align}
  \label{eq:SIS_E}
  \Stoich &= \left[ \begin{array}{rr}
      -1 & 1 \\
      1  & -1
    \end{array} \right], \quad
  R(x,y) = [\upsilon x_{1}y_{1},\gamma x_{2}]^{T}, \quad
  f(x,y) = \alpha x_{2}(x_{1}+x_{2})^{-1} -\beta y_{1}.
\end{align}
The rationale here is that infected individuals are recruited from the
susceptible population at a rate proportional to the concentration of
infectious substance. The concentration variable in turn is increased
at a rate proportional to the proportion of infected individuals, and
decreased according to a decay parameter $\beta$. It is
straightforward to show by non-dimensionalization that $\alpha = 1$
may be selected \review{without losing} any generality \cite{siminf_Ch}. The
SIS$_E$ model is summarized schematically in
\figref{fig:modelSchematic}.

\subsection*{Dynamic networks}

We shall adopt a somewhat more specific notation to capture a dynamic
contact network driven by data. We assume $\Nnodes$ nodes in total and
consider the state matrix $[\X;\;\Y] \in [\Intdom_{+}^{\Nspecies
    \times \Nnodes};\;\Realdom^{\Ndet \times \Nnodes}]$. The
\emph{local} dynamics is modeled by $\Nnodes$ replicas of
\eqref{eq:JSDE_ODE},
\begin{align}
  \label{eq:local}
  \begin{array}{rcl}
    d\X^{(i)}(t) &=& \Stoich\fatmu^{(i)}(dt) \\
    \frac{d}{dt}\Y^{(i)}(t) &=& f(\X^{(i)},\Y^{(i)}) \\
  \end{array} \qquad i = 1,\ldots,\Nnodes.
\end{align}
Assuming a given undirected contact graph $\mathcal{G}$ each node $i$
is allowed to affect the state of the nodes in the \emph{connected
  components} $C(i)$. Similarly, node $i$ is itself affected by all
nodes $j$ such that $i \in C(j)$. The resulting network dynamics can
then be written as
\begin{align}
  \begin{array}{rcl}
    d\X^{(i)}(t) &=& -\sum_{j \in C(i)} \Connect\fatnu^{(i,j)}(dt)+
    \sum_{j; \, i \in C(j)} \Connect\fatnu^{(j,i)}(dt) \\
    \frac{d}{dt} \Y^{(i)}(t) &=& -\sum_{j \in C(i)} g(\X^{(i)},\Y^{(i)})+
    \sum_{j; \, i \in C(j)} g(\X^{(j)},\Y^{(j)})
  \end{array}
\end{align}
for some flow intensities $\fatnu$ and $g$. Although the counting
measure $\fatnu^{(i,j)}$ may depend on the state of the sending node
$i$, the model of the paper relies fully on externally scheduled
events. Also, the specific model in the paper does not include flow of
the concentration variable and hence $g \equiv 0$. Note that, in any
case, mass balance is respected by this form of internodal
connections.

Using superposition of the local and the global dynamics we obtain
\begin{align}
  \label{eq:master}
  \begin{array}{rcl}
    d\X^{(i)}(t) &=& \Stoich\fatmu^{(i)}(dt)-
    \sum_{j \in C(i)} \Connect\fatnu^{(i,j)}(dt)+
    \sum_{j; \, i \in C(j)} \Connect\fatnu^{(j,i)}(dt) \\
    \frac{d}{dt} \Y^{(i)}(t) &=& f(\X^{(i)},\Y^{(i)})-
    \sum_{j \in C(i)} g(\X^{(i)},\Y^{(i)})+
    \sum_{j; \, i \in C(j)} g(\X^{(j)},\Y^{(j)})
  \end{array}
\end{align}
for $i = 1,\ldots,\Nnodes$. The SIS$_E$ model employed in the paper is
defined by \eqref{eq:SIS_E}, and with $(\mathcal{G},\Connect,\fatnu)$
taken from actual animal transport data, and $g \equiv 0$ as
mentioned. Under the mentioned non-dimensionalization the model
parameters are therefore tentatively $[\upsilon,\gamma,\beta]$. Due to
the fact that data is collected from a country with rather large
North-South climate variations (namely Sweden), the decay $\beta$ of
infectious substance is further separated into seasons
$[\beta_1,\beta_2,\beta_3,\beta_4]$ for
$[$spring,summer,fall,winter$]$, and where we put $\beta_1 = \beta_3$.
Data from the Swedish Meteorological and Hydrological Institute (SMHI)
was used to determine the duration of the seasons depending on each
node's geographical location. The seasons were defined by SMHI given
the average temperature for the reference period 1961–-1990: (winter)
below 0\degree C, (spring) between 0 and 10\degree C, (summer) above
10\degree C, and (autumn) between 0 and 10\degree C \cite{SMHI}.

\subsection*{Time discretization and simulation in parallel}

In order to simulate the dynamics effectively, time has to be
discretized. This is particularly reasonable in the current context
since data itself is sampled at a finite resolution. Put $0 = t_{0} <
t_{1} < t_2 < \cdots$ and define $\Delta t_n = t_{n+1}-t_n$. The
SIS$_E$ model in the paper is simulated by operator
splitting. Specifically, we employ the three-step method (recall that
$g \equiv 0$)
\begin{align}
  \label{eq:numstep3}
  \tilde{\X}_{n+1}^{(i)} &= \X_{n}^{(i)} + \int^{t_{n+1}}_{t_n} \Stoich
  \fatmu^{(i)}(\tilde{\X}^{(i)}(s),\Y_{n}^{(i)}; \; ds), \\
  \label{eq:numstep4}
  \X_{n+1}^{(i)} &= \tilde{\X}^{(i)}_{n+1}-\int^{t_{n+1}}_{t_n} \sum_{j \in C(i)}
  \Connect\fatnu^{(i,j)}(\X^{(i)}(s),\Y_{n}^{(i)}; \; ds)+
  \int^{t_{n+1}}_{t_n} \sum_{j; \, i \in C(j)}
  \Connect\fatnu^{(j,i)}(\X^{(i)}(s),\Y_{n}^{(i)}; \; ds), \\
  \label{eq:numstep5}
  \Y_{n+1}^{(i)} &= \Y_{n}^{(i)} + f(\tilde{\X}_{n+1}^{(i)},
  \Y_{n}^{(i)}) \, \Delta t_{n}.
\end{align}
The local stochastic dynamics is first simulated in time in
\eqref{eq:numstep3} to produce the temporary variable
$\tilde{\X}$. All connecting transport events are next incorporated in
\eqref{eq:numstep4}. Finally, in \eqref{eq:numstep5} the local
dynamics for the concentration variables is computed by the standard
Euler forward method in time with time-step $\Delta t_{n}$.


\section{Bayesian methodology}

This section discusses some additional details of the Bayesian
computational methodology developed in the paper.

\paragraph{Notation.} In the previous section we used a detailed
notation for the epidemiological process defined over $\Nnodes$
nodes. This level of detail is unnecessary here and we shall simply
write $X_t$ for this process.

\subsection*{Unfiltered and binary filtered data}

Recall that the epidemiological model is a stochastic process
$X_t = X_t(\theta)$, with parameter $\theta$. Define the prevalence
operator for a time series of state observations $(X_i) = (X(t_i))$ by
\begin{align}
  \Prev((X_i)) &= \left(\sum_j I_{ij}/\sum_j (S_{ij}+I_{ij})\right),
\end{align}
that is, $\Prev(\cdot)$ computes the total prevalence in the parts of
the network selected for observation and at the given points in
time. The resulting time series was what was considered as
``unfiltered data'' in the paper, remindful of the fact that the
prevalence is a deterministic function of the epidemiological
state. The part of the network selected for observation was the same
as the nodes where real data was available, thus facilitating a
comparison.

Define further the swab operator $\Swab((X_i))$ as the outcome of the
empirical stochastic swab protocol in \figref{fig:urnmodel}, at each
point in time $t_i$ and at each node $j$ selected for testing. By
construction, this is a stochastic function of the state and this time
series was considered a realistic replication of the available data
$D$, which was simply $D_{ij} = $ the outcome of the swab procedure at
time $t_i$ and at selected node $j$.

Let finally $B(\Swab((X_i)))$ and, respectively, $B(D)$ be the
accumulated versions, obtained by summing up the swab results in time
periods of quarter years and in $N_{\text{clusters}}$ node clusters
(only 1 cluster was used in the paper). These summed-up results were
referred to as the ``binary filtered data'' in the paper. Since the
\review{swab operator loses} a fair amount of information of the
epidemiological state, this case is expected to be more challenging.

\begin{SCfigure}
  \centering
  \includegraphics[width=0.45\linewidth]{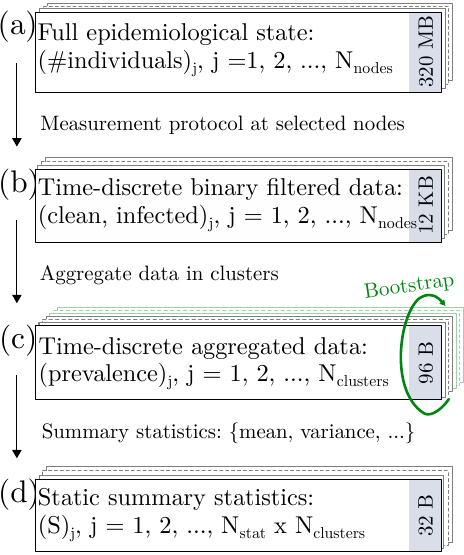}
  \caption{Data-flow from the model's state-space to the identifying
    summary statistics. {\bf (a)} Full state-space of the
    epidemiological model. {\bf (b)} After applying the pathogen
    detection protocol at selected nodes (see \figref{fig:urnmodel})
    data becomes time-discrete and binary. {\bf (c)} Aggregating the
    filtered state in clusters, and applying bootstrap resampling we
    obtain disease prevalence estimates at discrete points in time and
    in $N_{\text{clusters}}$ clusters (only 1 cluster was used in the
    paper). Finally, in {\bf (d)} static summary statistics are
    extracted. The actual size of the data in each step is indicated
    to the right; from beginning to end data shrinks 7 orders in
    magnitude.}
  \label{fig:dataToStat}
\end{SCfigure}

To conclude, in the notation of \algref{alg:ABC} and \algref{alg:SLAM}
below, and as used in our tests, the required simulator $F(\cdot)$ is
defined by $F(\theta) = \Prev((X_i))$ for unfiltered observations, and
$F(\theta) = B(\Swab((X_i)))$ for binary filtered data. The overall
processing of data collected from the \textit{in silico} disease model
is schematically summarized in \figref{fig:dataToStat}.

\subsection*{Summary statistics}

We based our Bayesian procedures on six summary statistics
coefficients, selected to capture seasonal dynamics as in
\cite{everitt2018bootstrapped}\cite{papamakarios2016fast}. The first
four statistics are weighted versions of the mean of the data in each
3-month season. The fifth and sixth summary statistics are the two
largest in magnitude Fourier coefficients of the data.  We compute and
normalize the weights $w_i$ for the $i$th statistics by
\begin{align}
  \label{eq:weights}
  \widetilde{w}_i &= \text{\# measurements used to compute } s_i, \\
  w_i &= \frac{\tilde{w}_i}{\sum_i \tilde{w}_i}.
\end{align}

\subsection*{ABC rejection with forward simulation}

In the paper, we use the uniform kernel function,
\begin{equation}
  K_\varepsilon =
  \begin{cases}
    1 & \text{if } ||s - s_{\text{obs}}|| < \varepsilon \\
    0 & \text{otherwise}
  \end{cases},
\end{equation}
and choose $\varepsilon$ such that only a fixed fraction of proposals
are accepted. The algorithm as we deploy it is summarized in
\algref{alg:ABC}.

\begin{algorithm}
  \caption{\textit{ABC rejection with forward simulation.}}
  \label{alg:ABC}
  \begin{algorithmic}
    \Require{Summarized data $s_{\text{obs}}$}
    \For{$i = 1,\dots,N_{\text{sample}}$}
    \Repeat{}
    \State{Propose $\theta^* \sim P(\theta)$}
    \State{Simulate $z \sim F(\theta^*)$}
    \State{Summarize $s = S(z)$}
        \Until{$K_\varepsilon(||s - s_{\text{obs}}||) = 1$}
    \State{$\theta_i = \theta^*$}
    \EndFor{}
  \end{algorithmic}
\end{algorithm}

\subsection*{Synthetic Likelihood Adaptive Metropolis with Empirical
  Bootstrapping}

We implemented the SLAM algorithm as described in the main paper in R
\cite{RSoftware}, and all the necessary data, computer codes, and
experimental scripts are available online at
\url{https://github.com/robineriksson/BayesianDataDrivenModeling}. An
in-depth description of the Bayesian sampling procedure is found in
\algref{alg:SLAM}.



\begin{algorithm} {\small
  \caption{\textit{SLAM with preprocessed Metropolized Independent Sampler.}}
  \label{alg:SLAM+}
  \begin{algorithmic}[1]
    \State{Generate $\theta_{1:N_{\text{train}}}$ using {\it SL within AM}}
    \State{Compute $\bar{C} = \text{Cov}(\theta_{1:N_{train}})$ and $\bar{\theta} = \text{Mean}(\theta_{1:N_{\text{train}}})$}
    \State{Generate $\theta_{1:N_{\text{sample}}}$ using {\it SL within MIS} with $\bar{C}$ and $\bar{\theta}$}
    \Function{SL}{$\theta$, $s_{\text{obs}}$}
    \State{Simulate
      $Y = \big( y_1, \dots, y_N \big), y_j \sim F(\theta)$}
    \State{Bootstrap
      $Z = \big( z_1, \dots, z_R \big), z_j \sim \hat{F}_N(Y)$}
    \State{Estimate
      $(\hat{\mu}_{\theta}, \widehat{\Sigma}_{\theta})$ from
      $S = \mathbf{S}(Z)$}
    \State{\Return{$\mathcal{L}_{\theta} = \Pr_{\theta}(s_{\text{obs}}|S)$}}
    \EndFunction{}
    \Function{SL within AM (``SLAM)}{$\theta_{1}$, $\mathcal{L}_{\theta_1}$}
    \For{$i = 2,\dots, N_{\text{train}}$}
    \State{$C_i :=$ \textbf{if} $i > i_0$ \textbf{then}
      $\xi \, \text{Cov}(\theta_1, \ldots ,\theta_{i-1}) + \xi \epsilon I$
      \textbf{else} $C_0$}
    \State{Sample
      $\theta^* \sim \mathcal{N}(\theta_{i-1},C_i)$}
    \State{Compute $L_{\theta^*} = \text{SL}(\theta^*)$}
    \State{$\theta_i = \text{accept/reject}(\theta_{i-1}, \theta^*, \mathcal{L}_\theta, \mathcal{L}_{\theta^*})$}
    \EndFor{}
    \State{\Return{$\theta_{1:N_{\text{\text{train}}}}$}}
    \EndFunction{}
    \Function{SL within MIS}{$\theta_0$, $\mathcal{L}_{\theta_0}$, $\bar{\theta}$, $\bar{C}$}
    \For{$i = 1,\dots, N_{\text{sample}}$}
    \State{Sample
      $\theta^* \sim \mathcal{N}(\bar{\theta}, \bar{C})$}
    \State{Compute $L_{\theta^*} = \text{SL}(\theta^*)$}
    \State{$\theta_i = \text{accept/reject}(\theta_{i-1}, \theta^*, \mathcal{L}_\theta, \mathcal{L}_{\theta^*})$}
    \EndFor{}
    \State{\Return{$\theta_{1:N_{\text{sample}}}$}}
    \EndFunction{}
    \Function{accept/reject}{$\theta_{i-1}$, $\theta^*$, $\mathcal{L}_\theta$, $\mathcal{L}_{\theta^*}$}
    \If{$\mathcal{U}(0,1) < \min \big( 1,\mathcal{L}_{\theta^*} /
      \mathcal{L}_\theta\big)$}
    \State{$\theta_i = \theta^*$ and
      $\mathcal{L}_\theta = \mathcal{L}_{\theta^{*}}$}
    \Else{}
    \State{$\theta_i = \theta_{i-1}$}
    \EndIf{}
    \State{\Return{$\theta_i$, $\mathcal{L}_\theta$}}
    \EndFunction{}
  \end{algorithmic}
  }
\end{algorithm}

\subsection*{Computational protocols and details}

Throughout this work we have emphasized the need for confidence in
each step forward: to achieve this we have followed the ``inverse
crime scheme''. Instead of starting at the final target set-up we
first test the proposed method on a less complex system where a
synthetic ground truth is used, and if the results are satisfactory we
add complexity. This procedure is repeated until the full system is
reached. On the one hand we reason at the negative side of things
here: there is clearly little hope in resolving a more complex set-up
without first handling an easier task. On the other hand, at each
successful iteration, one gains intuition, knowledge, and confidence
about the system and the method. For these reasons we first tested the
proposed method on a synthetic set-up at a small-scale 1600-nodes
version of the system and only then approached the full-scale version
in a second step. We also tested the method initially by assuming the
unfiltered state of the process could be measured, and only then moved
on to considering the more realistic noisy measurement filter.

\begin{figure}[h]
  \centering
  \includegraphics[angle=0, width = 1\linewidth]{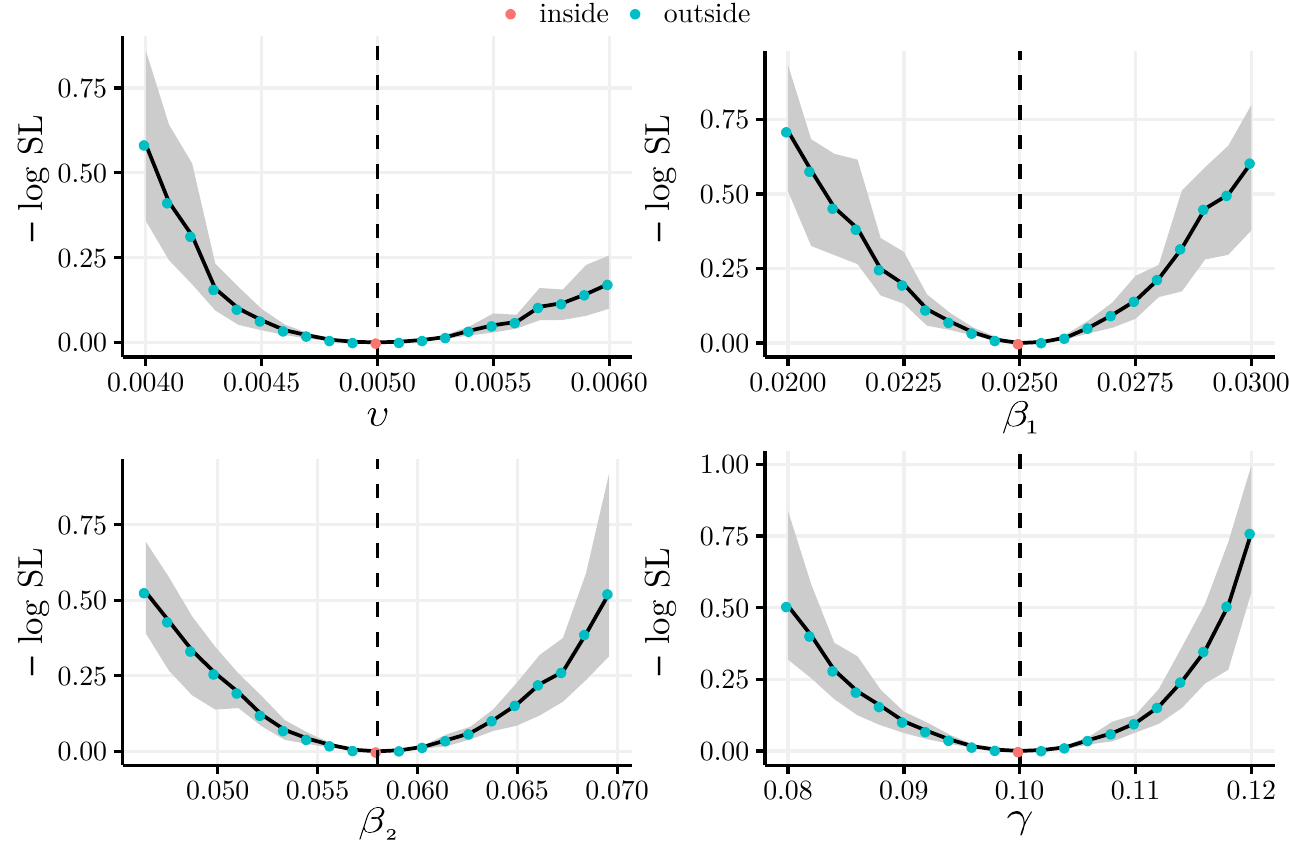}
  \caption{The response in $-\log \text{SL}$ when diverging from the
    true parameter values (dashed line) and with the other parameters
    held fixed at their true values,
    $(\upsilon, \beta_1, \beta_2, \gamma) = (0.005, 0.025,
    0.058,0.1)$. The SL was computed multiple $N = 20$ times, and the
    shaded gray area represents the 95\% CI. The red dots are either
    the minimum, or their CI overlaps with the minimum.  The mean
    coefficient of variation for the corresponding marginal
    distribution was $0.2\%$.}
  \label{fig:grid}
\end{figure}

As a very first initial test, we investigated the feasibility of the
parameterization by traversing a one-dimensional parameter grid on the
small-scale 1600-nodes four-parameter model.  We considered synthetic
data from a known truth, and we fixed all parameters except for
one. For this parameter we considered the domain
$\text{true value} \pm 20\%$ and estimated the synthetic likelihood
(SL) for the synthetic data, using multiple independent evaluations
for each point ($N = 20$). The purpose is to observe the definiteness
in any maxima of the SL for this parameter dimension alone. We did
this test for all parameter dimensions resulting in
\figref{fig:grid}. We found that $-\log \text{SL}$ is locally convex
in the (one-dimensional) vicinity of every parameter, and the maxima
are all apparently well-defined.

Next we moved on to considering all parameters at once, but still for
the smaller synthetic model. The inference methods tested here were
ABC rejection and SLAM, first using the unfiltered- and next the
binary filtered data. When comparing the two methods for the same type
of data, they were run for the same amount of time. For ABC, all
proposals were stored and afterwards we chose $\varepsilon$ in such a
way that the same amount of samples were selected as was produced by
SLAM. The results of the latter turned out to be considerably better
in our tests.

Finally we considered the full national scale model. In the
parameterization from actual measurements, we initialized the adaptive
MCMC chain with the point estimates suggested in
\cite{siminf3}. Because of the computationally costly character of the
problem, we connected $P = 10$ parallel chains on a cluster consisting
of Intel Xeon E5--2660 processors. We got $10 \times \numprint{1500}$
long chains from which we removed the first \numprint{500} elements as
burn-in, and considered these samples as the training set
$\theta_{1:N_{\text{train}}}$. The training set was then used in the
MIS procedure, which resulted in $10 \times \numprint{130}$ long
chains, after the removal of burn-in ($100$ per chain) and thinning ($10$).

For the AM specific parameters we used $\xi_d = 10^{-3}$,
$\epsilon_d = 10^{-5}$, $i_0 = 1$, $C_0 = 10^{-9} I_d$, where $i_0$ is
the number of Metropolis iterations for which $C_0$ is used as the
covariance matrix in the proposals (cf.~\algref{alg:SLAM}).

\paragraph{Results} The errors computed and referred to in the paper
are detailed in \tabref{tab:1600MultiSmall} and
\tabref{tab:RealMultiFull}. To be able to easily compare numbers we
report normalized root-mean-square errors
$\text{NRMSE} := \sqrt{\text{MSE}}/|\text{parameter value}|$.

\begin{figure}[htp]
  \centering
  \includegraphics[width=1\linewidth]{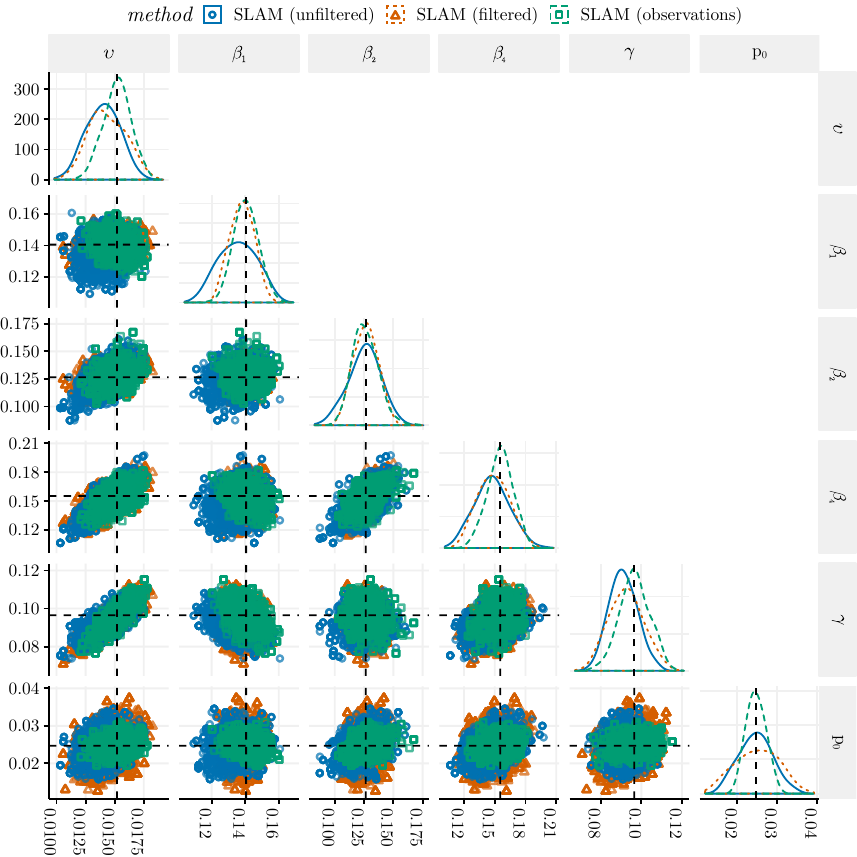}
  \caption{Posterior samples obtained by using SLAM on unfiltered-,
    binary filtered state, and real observations on the full
    data-driven simulation over \numprint{37220} nodes. The samples
    were obtained from \numprint{14000} accepted Metropolis steps,
    removing \numprint{1000} transient steps and thinning by 10. The
    values of the MMSE estimators thus obtained are for the case of
    real observations
    $(\upsilon, \beta_{1} [= \beta_{3}], \beta_{2}, \beta_{4}, \gamma,
    p_0) = (0.0151, 0.141, 0.126, 0.156, 0.0965, 0.0247)$.}
  \label{fig:AllRealMultiFull}
\end{figure}

Marginal distributions are shown in \figref{fig:AllRealMultiFull},
which differs from the one in the paper as it contains all the
estimated parameters.

\begin{table}[htp]
  \centering
  \caption{Errors of the ABC- and the SLAM-estimators for the
    synthetic small-scale problem. Normalized root-mean-square errors
    (NRMSE) of the minimum mean square error (MMSE) parameter
    estimators.}
  \label{tab:1600MultiSmall}
  \begin{tabular}{l r r r r}
    \toprule
    & ABC (Unfilt) & ABC (Filt) & SLAM (Unfilt) & SLAM (Filt) \\
    \midrule
    $\upsilon$ &$0.42$ & $0.42$ & $0.11$ & $0.042$ \\
    $\beta_{1}$ &$0.45$ & $0.41$ & $0.23$ & $0.078$ \\
    $\beta_{2}$ & $0.31$ & $0.40$ & $0.12$ & $0.021$ \\
    $\gamma$ & $0.42$ & $0.39$ & $0.04$ & $0.031$ \\
    \bottomrule
  \end{tabular}
\end{table}

\begin{table}[htp]
  \centering
  \caption{Errors of SLAM for the full problem. NRMSE with imputed
    bias determined from 140 accepted posterior samples. See also
    \figref{fig:AllRealMultiFull}.}
  \label{tab:RealMultiFull}
  \begin{tabular}{l r  r  r  r }
    \toprule
    Param
    & SLAM (Unfilt)
    & SLAM (Filt)
    & SLAM (Obs) [variance]
    & SLAM (Obs) [NMRSE]\\
    \midrule
    $\upsilon$ & $0.11$ & $0.12$ & $1.24\cdot 10^{-6}$ & $0.08$ \\
    $\beta_{1} [= \beta_{3}]$ & $0.05$ & $0.09$ & $4.87 \cdot 10^{-5}$ & $0.05$ \\
    $\beta_{2}$ & $0.09 $ & $0.12$ & $1.11 \cdot 10^{-4}$ & $0.09$ \\
    $\beta_{4}$ & $0.10$ & $0.11$ & $1.37 \cdot 10^{-4}$ & $0.08$ \\
    $\gamma$    & $0.09$ & $0.08$ & $5.00 \cdot 10^{-5}$ & $0.08$ \\
    $p_0$       & $0.22$ & $0.17$ & $5.49 \cdot 10^{-6}$ & $0.11$ \\
    \bottomrule
  \end{tabular}
\end{table}

\begin{table}[htp]
  \centering
  \caption{Gelman-Rubin diagnostic on each parameter in the Markov
    chains generated from using SLAM on unfiltered-, binary filtered
    state, and real observations on the full data-driven simulation
    over \numprint{37220} nodes}
  \label{tab:gelman}
  \begin{tabular}{l r  r  r  r }
    \toprule
    Param
    & SLAM (Unfilt)
    & SLAM (Filt)
    & SLAM (Obs) \\  \midrule
    $\upsilon$ & $1.032$ & $1.025$ & $1.050$  \\
    $\beta_{1} [= \beta_{3}]$ & $1.024$ & $1.008$ & $1.044$ \\
    $\beta_{2}$ & $1.020 $ & $1.060$ &  $1.022$ \\
    $\beta_{4}$ & $1.031$ & $1.061$ & $1.017$ \\
    $\gamma$    & $1.030$ & $1.017$ & $1.063$ \\
    $p_0$       & $1.036$ & $1.031$ & $1.038$ \\
    \bottomrule
  \end{tabular}
\end{table}

\subsection*{Intervention evaluation}

In the paper, we evaluated three proposed intervention techniques.
Besides the graphical investigation in the main paper we also
quantified the reduction factor of the interventions, see
\tabref{tab:intervention}. In the table, the three interventions are
measured at two points in time, and are evaluated against the case
with no intervention. The quantities in the table are means with 95\%
CI. From these we can cannot separate intervention \#3 and \#4 from
each other as their intervals overlap. However, they are both
significantly better than intervention \#2.

\begin{table}[htp]
  \caption{Reduction factors (mean and 95\% CI) for the different
    intervention strategies.}
  \label{tab:intervention}
  \centering
  \begin{tabular}{l r  r }
    \toprule
    Intervention strategy & Year 1 & Year 3 \\
    \midrule
    2) No transmission via transport & $0.90 \, [0.87, 0.93]$ &
    $0.79 \, [0.74, 0.83]$ \\
    3) 10\% increased decay $\beta$ & $0.33 \, [0.30, 0.36]$ &
    $0.089 \, [0.072, 0.11]$ \\
    4) 10\% reduced uptake $\upsilon$ & $0.30 \, [0.26, 0.32]$ &
    $0.068 \, [0.053, 0.089]$ \\
    \bottomrule
  \end{tabular}
\end{table}


\newcommand{\available}[1]{Available at \texttt{#1}}


\end{document}